\documentclass[usenatbib]{mn2e}
\input{psfig.sty}
\usepackage{graphicx}

\def\beq{\begin{equation}}
\def\enq{\end{equation}}

\title{Spin and Spectral Variations of Peculiar High-Mass X-ray Binary 4U 2206+54}

\author[Wei Wang]{Wei Wang\thanks{E-mail: wangwei@bao.ac.cn} \\
National Astronomical Observatories, Chinese Academy of Sciences,
Beijing 100012, China}

\begin{document}
\maketitle

\begin{abstract}
Spin properties and spectral variations of high mass X-ray binary 4U 2206+54 are studied with long-term hard X-ray monitoring observations by INTEGRAL. A long-period X-ray pulsar of $P_{\rm spin}\sim 5558$ s has been identified in 4U 2206+54. The spin evolution of the neutron star in 4U 2206+54 is detected with the INTEGRAL/IBIS data. From 2005 to 2011, the spin period of the neutron star in 4U 2206+54 varies from $\sim$ 5558 s to $\sim$ 5588 s. The average spin-down rate in the last 20 years is derived as $\sim 5\times 10^{-7}$ s s$^{-1}$. 4U 2206+54 is a variable source with luminosities of $\sim 10^{35} - 10^{36}$ erg s$^{-1}$ in the range of 3 -- 100 keV. Its spectrum can be described by an absorbed power-law model with exponential rolloff. The hydrogen column density and photon index show the anti-correlations with hard X-ray luminosity: low column density and small photon index at maximum of luminosity. This spectral variation pattern suggests that 4U 2206+54 would be a highly obscured binary system. Furthermore, the possible cyclotron line features are searched for in the spectra of different luminosities. The possible cyclotron absorption feature around 30 keV is not confirmed in our work. The physical origin and evolutional track of this very slow pulsation neutron star are still unclear. We suggest that 4U 2206+54 would be a young system with the long-term spin-down trend, would evolve to a longer spin period range and transit to the spin-up phase similar to 2S 0114+65. These very slow pulsation X-ray pulsars would belong to a new class of compact objects -- accreting magnetars.
\end{abstract}

\begin{keywords} stars: individual (4U 2206+54) -- stars: neutron -- stars: magnetic
fields -- stars : binaries : close -- X-rays: binaries.
\end{keywords}

\section{Introduction}

4U 2206+54 is a peculiar high mass X-ray binary, emitting the persistent X-rays with a main sequence star companion (Negueruela \& Reig 2001; Rib\'o et al. 2006;  Blay et al. 2006). Its X-ray luminosity varies in the range of $L_x\sim 10^{34}-10^{36}$ erg s$^{-1}$ from 1 -- 10 keV according to the RXTE, BeppoSAX, SWIFT and INTEGRAL light curves between 1996 and 2007 (Torrejon et al. 2004; Masetti et al. 2004; Blay et al. 2005; Corbet et al. 2007; Wang 2009, 2010). X-ray monitoring by RXTE has suggested an orbital period of 9.6 days in 4U 2206+54(Corbet \& Peele 2001). Recently SWIFT/BAT observations (Corbet et al. 2007) and
RXTE/ASM data (Wang 2009; Levine et al. 2010) found a modulation of $\sim 19.12$ days, twice of the 9.6-day period. Then this 19.12-day period was thought to be the true orbital period of 4U 2206+54 (Corbet et al. 2007; Wang 2009).

The nature of the compact object in 4U 2206+54 is also in dispute
for a long time (Negueruela \& Reig 2001; Corbet \& Peele 2001). Recent reports on the possible detection of electron cyclotron resonant absorption line features at $\sim 30$ keV and $\sim 60$ keV suggest a magnetic field of $\sim 3\times 10^{12}$ G by different
observations of RXTE, BeppoSAX and INTEGRAL (Torrejon et al. 2004;
Masetti et al. 2004; Blay et al. 2005; Wang 2009). The discovery of
a 5560 s pulsation in light curves of 4U 2206+54 from
RXTE, INTEGRAL and Suzaku observations suggests that it would be
a X-ray pulsar (Reig et al. 2009; Wang 2009; Finger et al. 2010).

\begin{table*}
\scriptsize
\caption{\scriptsize INTEGRAL/IBIS observations of the field around 4U
2206+54 within the off-axis angle below 10$^\circ$. The time intervals of observations in the revolution
number and the corresponding MJD dates, the corrected on-source
exposure time, and the average off-axis angle on the source are listed. Mean IBIS count rate and the detection
significance level value in the energy range of 20 -- 80 keV were
also shown.}
\begin{center}
\begin{tabular}{l c c c c l}
\hline \hline
Rev. Num. &  Obs Date (MJD)  &  IBIS rate  & Detection significance level &  On-source time (ks) & off-axis angle\\
\hline
510 & 54085.9--54088.1  &  5.14 $\pm$ 0.08 & 65$\sigma$  &   176 & 1.9$^\circ$ \\
511 & 54088.8--54091.2  &  1.89 $\pm$ 0.09 & 20$\sigma$  &   191 & 1.3$^\circ$ \\
512 & 54091.9--54094.1  &  2.50 $\pm$ 0.09 & 27$\sigma$  &   196 & 1.5$^\circ$ \\
513 & 54094.8--54097.3  &  0.91 $\pm$ 0.09 & 10$\sigma$  &   165 & 2.5$^\circ$ \\
514 & 54097.8--54098.7  &  1.80 $\pm$ 0.11  & 16$\sigma$  &   68 & 3.1$^\circ$  \\
515 & 54100.9--54101.7  &  $<0.6$      & $<5\sigma$  &   61  & 3.5$^\circ$ \\
516 & 54103.8--54104.6  &  2.04 $\pm$ 0.13 & 16$\sigma$  &   64 & 3.4$^\circ$ \\
517 & 54106.7--54107.2  &  1.37 $\pm$ 0.09 & 14$\sigma$  &   37 & 3.8$^\circ$  \\
518 & 54109.7--54110.1  &  1.48 $\pm$ 0.14  & 10$\sigma$  &   34  & 3.2$^\circ$ \\
519 & 54112.7--54113.2  &  2.53 $\pm$ 0.22  & 12$\sigma$  &   38 & 3.1$^\circ$  \\
520 & 54115.8--54116.2  &  2.76 $\pm$ 0.23  & 11$\sigma$  &   36  & 2.5$^\circ$ \\
553 & 54215.4--54217.0  &  0.81 $\pm$ 0.11  & 8 $\sigma$  &   117 & 3.2$^\circ$ \\
557 & 54226.7--54229.0  &  1.12 $\pm$ 0.09 & 12$\sigma$  &   171 & 1.8$^\circ$  \\
558 & 54230.9--54231.9  &  1.70 $\pm$ 0.12  & 14$\sigma$  &   77  & 1.9$^\circ$ \\
559 & 54232.7--54235.0  &  1.68 $\pm$ 0.09 & 18$\sigma$  &   163 & 1.4$^\circ$ \\
562 & 54241.4--54241.6  &  1.52 $\pm$ 0.30  & 5 $\sigma$  &   18 & 2.7$^\circ$  \\
563 & 54245.3--54246.9  &  1.02 $\pm$ 0.11  & 9 $\sigma$  &   123 & 2.1$^\circ$ \\
564 & 54247.4--54249.9  &  1.28 $\pm$ 0.08 & 15$\sigma$  &   189 & 1.4$^\circ$ \\
567 & 54256.3--54258.9  &  2.72 $\pm$ 0.08 & 34$\sigma$  &   204 & 1.2$^\circ$ \\
568 & 54259.4--54261.4  &  1.74 $\pm$ 0.08 & 21$\sigma$  &   203 & 1.4$^\circ$ \\
569 & 54262.4--54263.1  &  3.13 $\pm$ 0.21  & 15$\sigma$  &   50  & 1.8$^\circ$ \\
583 & 54304.2--54304.5  &  1.44 $\pm$ 0.21  & 7 $\sigma$  &   22  & 2.9$^\circ$ \\
649 & 54401.6--54404.0  &  $<0.6$      & $<5\sigma$  &   57  & 3.6$^\circ$ \\
652 & 54410.6--54412.9  &  $<0.6$      & $<5\sigma$  &   56  & 4.1$^\circ$ \\
998 & 55546.9--55547.6  &  $2.41\pm 0.20$ & 12$\sigma$ &  50 & 7.1$^\circ$ \\
1007 & 55572.6--55575.2 &  $3.58\pm 0.19$ & 19$\sigma$ & 81 & 8.5$^\circ$\\
1008 & 55576.6--55577.3 &  $3.74\pm 0.19$ & 20$\sigma$  & 52 & 6.5$^\circ$ \\
1048 & 55695.6--55695.9 &  $1.95\pm 0.23$ & 8$\sigma$ & 19 & 7.0$^\circ$\\
1049 & 55700.2--55700.6 & $1.85\pm 0.25$ & 7$\sigma$ &  19 & 7.8$^\circ$\\
1050 & 55702.4--55703.3 & $<1.1$ &  $<5\sigma$ & 25 & 8.9$^\circ$ \\
1051 & 55704.3--55705.5 &  $2.70\pm 0.19$ & 14$\sigma$ & 24 & 6.2$^\circ$\\
1066 & 55709.6--55710.2 & $2.11\pm 0.21$ & 10$\sigma$ & 35 & 6.6$^\circ$\\
1072 & 55769.0--55769.6 & $3.53\pm 0.21$ & 17$\sigma$ & 40  & 7.0$^\circ$\\
1075 & 55776.1--55777.1 & $1.52\pm 0.19$ & 8$\sigma$ & 52 & 6.7$^\circ$\\
1076 & 55779.1--55781.5 & $<1.0$ & $<5\sigma$ &  30 & 8.8$^\circ$\\
\hline

\end{tabular}
\end{center}
\end{table*}

\begin{figure*}
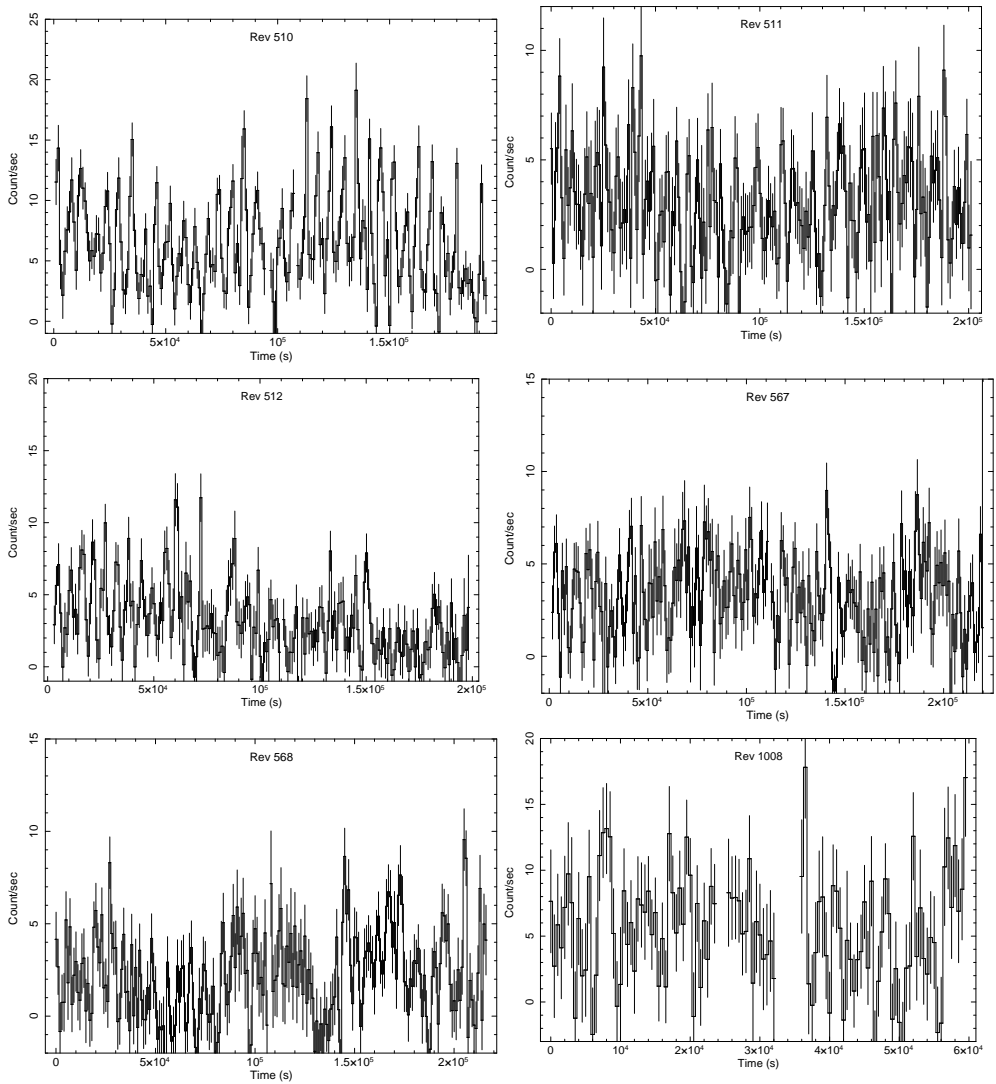

\centering
\includegraphics[angle=-90,width=6.5cm]{4u2206_lc510.ps}
\includegraphics[angle=-90,width=6.5cm]{4u2206_lc511.ps}
\includegraphics[angle=-90,width=6.5cm]{4u2206_lc512.ps}
\includegraphics[angle=-90,width=6.5cm]{4u2206_lc567.ps}
\includegraphics[angle=-90,width=6.5cm]{4u2206_lc568.ps}
\includegraphics[angle=-90,width=6.5cm]{4u2206_lc1008.eps}
\caption{Background-subtracted hard X-ray light curves of 4U 2206+54 in the energy band of 20 -- 80 keV
for six revolutions in which 4U 2206+54 is detected by IBIS with significance level $>20\sigma$. }
\end{figure*}

\begin{figure*}
\centering
\includegraphics[angle=0,width=13cm]{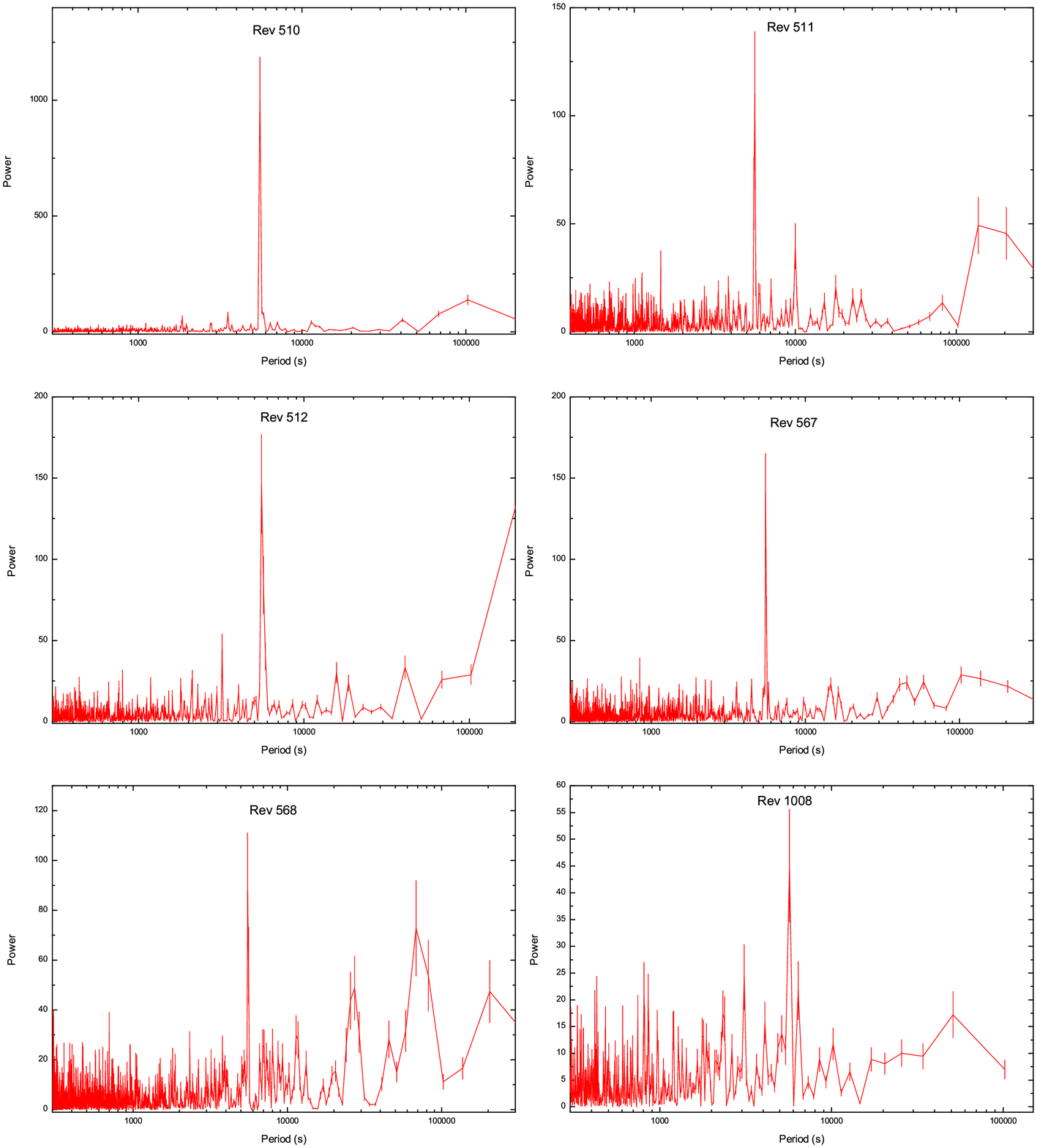}
\caption
{The power spectra of the hard X-ray light curves in the range of 20 -- 80 keV for the six revolutions in Fig. 1. The significant peaks at $\sim 5500$ s appear in all power spectra.}
\end{figure*}

The physical origin of variability pattern and slow-pulsation period (longer than 5000 s) is still unclear. Thus, the long-term X-ray monitoring on the source to study the spectral variability and spin properties is necessary. Finger et al. (2010) found a possible spin-down trend of 4U 2206+54 when they compared two spin period values derived by the Suzaku observations in 2007 and BeppoSAX detections in 1998. If the neutron star in 4U 2206+54 still spins down, this slow pulsation neutron star might become the slowest pulsation one of our known sources within several hundred years. Now INTEGRAL has collected more than eight years of observational data, and specially has done several pointing observations on 4U 2206+54 with a long exposure time ($>100$ ks) in each orbital revolution (3 days). The long on-source observations are very helpful to determine the spin period of the slow-pulsation neutron star. With repeated pointing observations on 4U 2206+54 with INTEGRAL, we try to study spin properties of the slow-pulsation neutron star, and determine the possible spin-down or spin-up of the neutron star in our observed epoch. Furthermore, we also use the long-term monitoring data of INTEGRAL to probe the spectral property variations of 4U 2206+54 with different accretion luminosities. These studies will help us to understand physical origin of variability pattern and formation scenario of slow-pulsation period.

This paper will be organized as follows. We will first introduce the INTEGRAL observations and data analysis in \S 2. In \S 3, spin properties of 4U 2206+54 are studied, and try to search for the spin derivative of 4U 2206+54. In \S 4, spectral properties and variations of 4U 2206+54 in different accretion states are studied. In addition, we try to search for the possible cyclotron absorption line features in hard X-ray spectrum of 4U 2206+54 which were reported in previous work. The summary and discussion are presented in \S 5.

\section{INTEGRAL Observations}

INTEGRAL is an ESA's currently operational space-based
hard X-ray/soft gamma-ray telescope (Winkler et al. 2003). There are two main
instruments aboard INTEGRAL, the imager IBIS (Ubertini et al.
2003) and the spectrometer SPI (Vedrenne et al. 2003),
supplemented by two X-ray monitors JEM-X (Lund et al. 2003) and an
optical monitor OMC (Mas-Hesse et al. 2003). All four instruments
are co-aligned, allowing simultaneous observations in a wide
energy range.

SPI has a lower sensitivity than IBIS for weak
sources below 200 keV, and detection of weak point sources is not
so significant for spectral studies. 4U 2206+54 is not detected by SPI. Then we will mainly use the data
collected with the low-energy array called IBIS-ISGRI (INTEGRAL
Soft Gamma-Ray Imager, Lebrun et al. 2003). IBIS/ISGRI has a 12' (FWHM)
angular resolution and arcmin source location accuracy in the
energy band of 15 -- 200 keV. JEM-X as the small X-ray detector collects the lower energy
photons from 3 -- 35 keV which is used to constrain the soft X-ray band spectral properties of 4U 2206+54 combined with IBIS.

4U 2206+54 is observed frequently during INTEGRAL all-sky surveys. In this
work, we use the available archival data for the IBIS observations
where 4U 2206+54 is within $\sim 10$ degrees of the pointing
direction with the on-source time longer than 15 ks. In Table 1, the information of available INTEGRAL observations used in this paper is summarized. The INTEGRAL observations monitored 4U 2206+54 from Dec 2006 -- Aug 2011. The archival data used in our work are available from
the INTEGRAL Science Data Center (ISDC).

The analysis is done with the standard INTEGRAL off-line
scientific analysis (OSA, Goldwurn et al. 2003) software, ver.
9.0. Individual pointings in each satellite revolution (3 days) processed with OSA 9.0 are mosaicked to create sky images for the source detection. We have used the energy range of 20 -- 80 keV by IBIS for source detection and quoted source fluxes for each revolution (see
Table 1). If 4U 2206+54 is detected by IBIS with significance level below 5$\sigma$, only upper limits of the count rate are given.


\section{Spin period of the neutron star in 4U 2206+54}

The spin period of the neutron star in 4U 2206+54 was discovered in 2009 (Reig et al. 2009; Wang 2009). Since 4U 2206+54 is a weak X-ray source and the continuous observing time is relative very short (compared with the time scale of the rotation period, $\sim$ 5500 s), uncertainties of determining the spin period and its possible derivative of 4U 2206+54 still exist. Reig et al. (2009) found the spin period of 5559$\pm 3$ s around MJD 54240 using RXTE/PCA data; and obtained a spin period of 5525$\pm 30$ s using EXOSAT observations in three occasions: 8 August 1983, 7 December 1984 and 27 June 1985. Wang (2009, 2010) found a spin period of $\sim 5550\pm 50$ around MJD 53719 using INTEGRAL/IBIS data. Finger et al. (2010) obtained a spin period of 5554$\pm 9$ s at MJD 54237 using the Suzaku data; and they also found a spin-period of 5420$\pm 28$ s around MJD 51141 by re-analyzing the BeppoSAX data, suggesting a spin down of the neutron star in 4U 2206+54 with a spin-down rate of $\sim (5.0\pm 1.3)\times 10^{-7}$ s s$^{-1}$ when they compared the results by Suzaku and BeppoSAX. Recently, Reig et al. (2012) reported a spin period around $5590\pm 10$ s at MJD 55599 using XMM-Newton observations. This result implies that the long-term spin-down trend of the neutron star in 4U 2206+54 is still going on. Spin property specially the spin period evolution of this slow pulsation X-ray pulsar is a key question to probe the physical nature of the object, still requiring more detailed studies.

INTEGRAL has a high Earth orbit with a 3 day period, which should have advantage to study the spin properties of this slow-pulsation neutron star. Recent INTEGRAL pointing observations on 4U 2206+54 have several revolutions with a long on-source time near 200 ks (see Table 1), these observations will provide us a chance to re-analyzed the spin-period of 4U 2206+54. In Fig. 1, the hard X-ray light curves are displayed in six revolutions when 4U 2206+54 was detected by IBIS observations with a significance level higher than $\sim$ 20 $\sigma$: Revs 510, 511, 512, 567, 568 and 1008. The background-subtracted light curves after barycentric corrections for the six revolutions in the range of 20 -- 80 keV are obtained with the time resolution of $\sim 50$ s (in Fig. 1, data points have been re-binned to 1000 s for clarity). The modulations in light curves are seen, specially for Rev 510 when 4U 2206+54 was detected with the highest significance level in our analysis.

Power spectra for the six light curves are presented in Fig. 2. The significant peak at $\sim 5500$ s is clearly detected in the all power spectra, suggesting a long spin period the neutron star located in 4U 2206+54. The detailed studies on the power spectrum of each revolution find that the peak at $\sim 5560$ s for five revolutions 510, 511, 512, 567 and 568; but for Rev 1008, the detected peak is around 5590 s. For Rev 510, the periodical signal in hard X-ray light curve is strongest in our observations, and we use the {\it efsearch} method to derive the spin period of 4U 2206+54 at $\sim 5558\pm 2$ s. Combining with {\it efsearch} and the pulse-folding technique, we also obtain the spin period values for other five revolutions (see Table 2): 5560$\pm 10$ s for Rev 511, 5560$\pm 7$ s for Rev 512, $5564\pm 7$ s for Rev 567, $5563\pm 10$ s for Rev 568 and $5588\pm 10$ s for Rev 1008. According to the IBIS results on the spin period values from 2005 -- 2011, 4U 2206+54 would undergo a long-term spin-down trend. Finally, we collect the previous spin period measurements including the present work in Fig. 3. Comparing the spin-period values derived in the last nearly twenty years, the reported spin-down trend of the neutron star in 4U 2206+54 is confirmed with a mean spin-down rate of $\sim 5\times 10^{-7}$ s s$^{-1}$.

\begin{figure}
\centering
\includegraphics[angle=0,width=9cm]{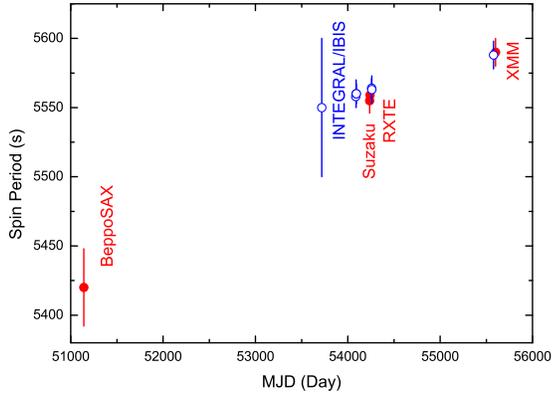}
\caption
{The spin period of the neutron star in 4U 2206+54 determined by different measurements (data points from Finger et al. 2010; Reig et al. 2009, 2012; Wang 2010 and this work). }

\end{figure}

\begin{figure}
\centering
\includegraphics[angle=0,width=9cm]{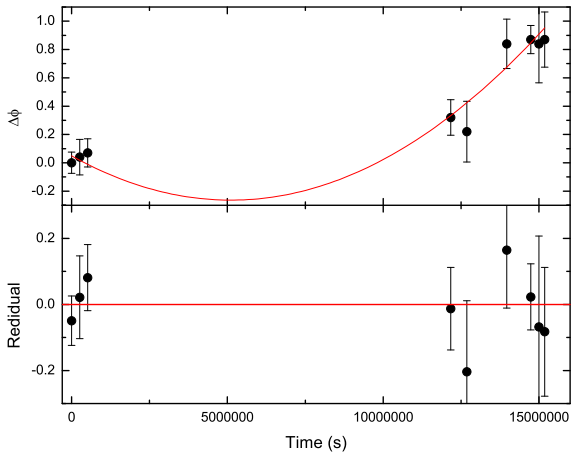}
\caption
{The pulse peak phase shifts obtained using a constant folding period (5560 s) with the best-fit second-order polynomial overplotted (upper panel) and the residual plots obtained with the folding parameters $P_{\rm spin}\sim 5557.1$ s and $\dot P_{\rm spin}\sim 6\times 10^{-7}$ s s$^{-1}$ (lower panel). The time scale is in seconds after the reference time $t_0$ which is the central time of Rev 510. }

\end{figure}

\begin{table}
 \caption{The spin period of 4U 2206+54 in six observational revolutions}
\begin{center}
\begin{tabular}{l c l}
\hline \hline
Rev No.  &   MJD &  Spin Period (s) \\
\hline
510 & 54087  & $5558\pm 2$ \\
511 & 54090  &  5560$\pm 10$   \\
512 & 54093  &   5560$\pm 7$ \\
567 &  54258 &  5564$\pm 7$ \\
568 & 54260 &  5563$\pm 10$ \\
1008 & 55577  &  $5588\pm 10$   \\
\hline
\end{tabular}
\end{center}

\end{table}

\begin{figure}
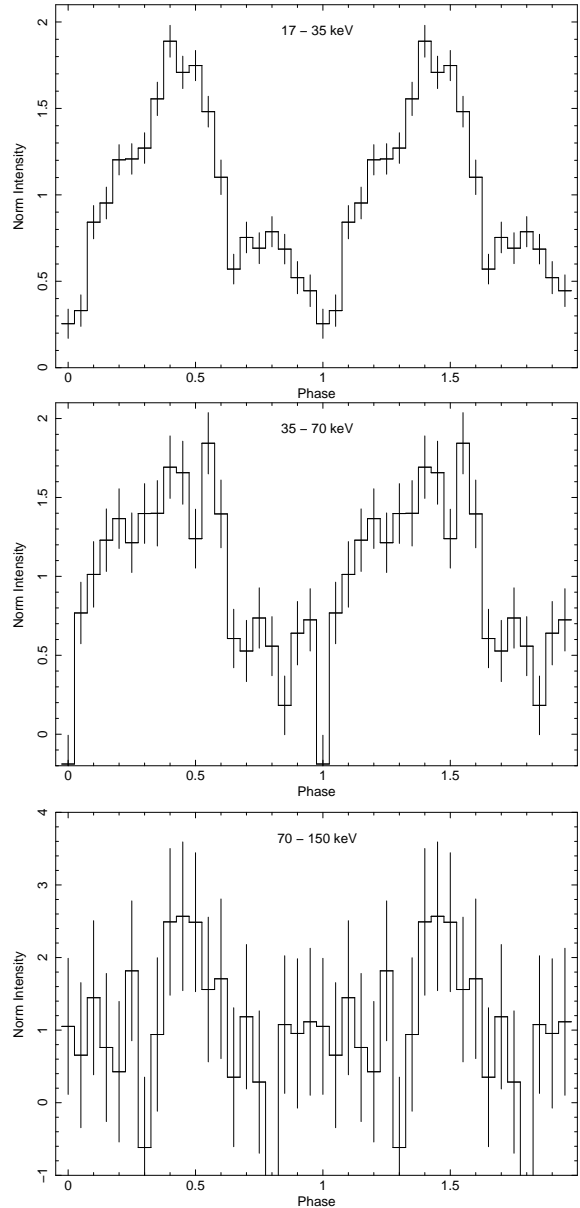

\centering
\includegraphics[angle=-90,width=7.5cm]{2206_510_17.eps}
\includegraphics[angle=-90,width=7.5cm]{2206_510_35.eps}
\includegraphics[angle=-90,width=7.5cm]{2206_510_70.eps}
\caption{The IBIS/ISGRI background subtracted light curves of 4U 2206+54 for Rev 510 folded at a pulsation period (5558 s) in three
different energy ranges: 17 -- 35 keV; 35 -- 70 keV; 70 -- 150 keV.  }

\end{figure}

The data points in Fig. 3 show the long-term spin-down trend of the neutron star in 4U 2206+54 from 1990s - 2011. However we also need to know more details of spin evolution of this object, i.e., in the relative short time scale, the spin-down rate is still observable or even possible the spin torque could be reversed. Therefore, to search for the possible spin evolution of the neutron star in a shorter time scale, we apply the phase shift technique in light curve analysis for 4U 2206+54 from Dec 2005 -- June 2006. We first obtain the power spectrum of the light curve for each revolution during the half year, and selected the revolutions in which the significant single peak around 5560 s ($>3 \sigma$) is detected in power spectra. We use light curve data of nine observational revolutions: 510, 511, 512, 557, 559 564, 567, 568, and 569. Then we fold all the hard X-ray light curves (background-subtracted and barycentric corrected with the 50 s time resolution) with a tentative period 5560 s to obtain the pulse profile of each revolution. We would calculate the phase ($\phi$) of the peak position for the pulse profile of each evolution (similar to methods in \S 3.2 of Ferrigno et al. 2007). We have used the phase of the peak in Rev 510 as the reference, and then derived the phase shift by $\Delta\phi(t)=\phi(t)-\phi(t_0)$, where $t_0$ is the central time of Rev 510, $t$ is the central time of other revolutions relative to $t_0$. Generally, we can reconstruct the actual phase shift by fitting $\Delta\phi(t)$  with a second-order polynomial \beq \Delta\phi(t)=\phi_0+at + bt^2, \enq where $\phi_0$ is the starting phase shift determined by the fit. The coefficient $a$ can be used to correct the tentative
choice of the folding frequency $\nu_t$ as \beq \nu_c=\nu_t-a, \enq where $\nu_c$ is the real pulse frequency at the reference time, i.e., Rev 510. The coefficient $b$ can be
used to find the possible pulse frequency derivative during the observational
period by the relation \beq \dot \nu=-2b. \enq

The calculated data points ($\Delta \phi$ versus time) for nine revolutions are plotted in Fig. 4. The second-order polynomial fitting is carried out and gives the pulse period of the neutron star of 4U 2206+54 in Rev. 510 at $5557.1\pm 1.9$ s which is consistent with the our previous result estimated with the {\it efsearch} method, and the average frequency derivative at $-(1.8\pm 0.5)\times 10^{-14}$ Hz s$^{-1}$ from Rev. 510 - Rev 569 with the reduced $\chi^2$ of 0.695 (6 d.o.f.). Though the significance level is not so high ($\sim 3\sigma$), the spin-down trend is still detected in the X-ray pulsar of 4U 2206+54 in the half year time scale (Dec 2005 -- June 2006), and the spin-down rate is also consistent with the average rate according to the long-term observations from 1990s -- 2011.

In Fig. 5, the folded pulse profiles at the determined spin period of 5558 s for Rev 510 are displayed. The hard X-ray light curves are presented in three energy ranges: 17 -- 35 keV; 35 -- 70 keV and 70 -- 150 keV. The pulse fraction (defined as ratio of the maximum of light curves minus the minimum over the maximum) is $\sim 86\%$ for the 17 -- 35 keV band, near $100\%$ for the 35 -- 70 keV and 70 -- 150 keV bands. The pulse profiles generally show a main peak around phase 0.4 -- 0.5; and the small second peak around phase 0.8 was also detected in the low energy band of 17 -- 35 keV.


\begin{figure*}
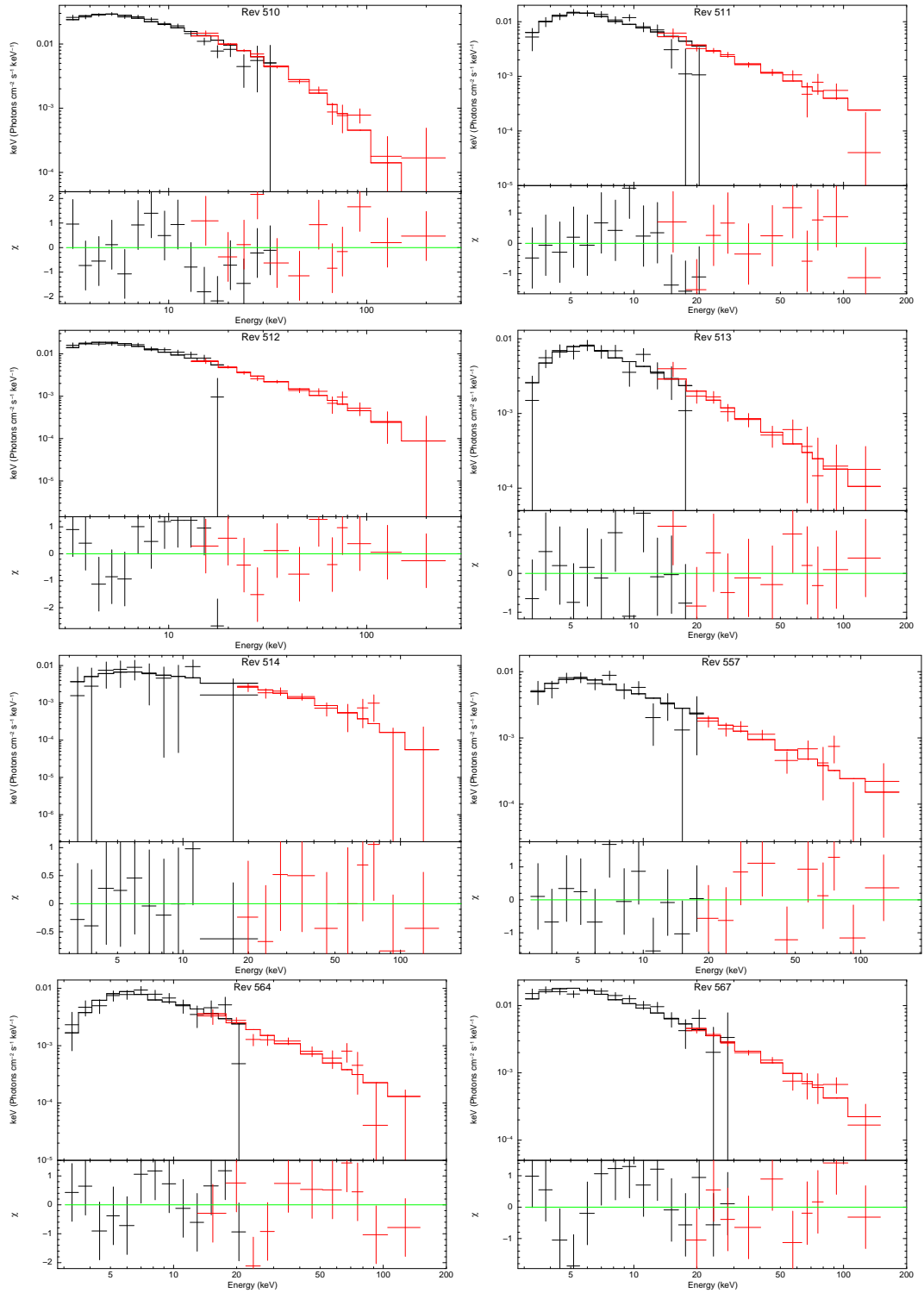

\centering
\includegraphics[angle=-90,width=7.0cm]{4u2206_cutpl510_euf.eps}
\includegraphics[angle=-90,width=7.0cm]{4u2206_cutpl511_euf.eps}
\includegraphics[angle=-90,width=7.0cm]{4u2206_cutpl512_euf.eps}
\includegraphics[angle=-90,width=7.0cm]{4u2206_cutpl513_euf.eps}
\includegraphics[angle=-90,width=7.0cm]{4u2206_cutpl514_euf.eps}
\includegraphics[angle=-90,width=7.0cm]{4u2206_cutpl557_euf.eps}
\includegraphics[angle=-90,width=7.0cm]{4u2206_cutpl564_euf.eps}
\includegraphics[angle=-90,width=7.0cm]{4u2206_cutpl567_euf.eps}
\caption{The hard X-ray spectra of 4U 2206+54 obtained by
JEMX and IBIS in eight different observational revolutions (also see Table 3). The spectra are fitted by an absorbed power-law plus high energy exponential roll-off ({\em abs*cutoffpl}).}

\end{figure*}

\section{Spectral properties and variations}

4U 2206+54 is a variable X-ray source whose spectral properties would vary with different luminosities. We will study the spectral properties of 4U 2206+54 in the energy range of 3 -- 200 keV combined with simultaneous observations by two detectors JEMX (3 -- 30 keV) and IBIS (18 -- 200 keV) aboard INTEGRAL. JEM-X has a small field of view so we can study 4U 2206+54 using
JEM-X only when the mean off-axis angle on the source is below
5$^\circ$ (see Table 1). The cross-calibration studies on the JEM-X and IBIS/ISGRI
detectors have been done by Jourdain et al. (2008) using the Crab
observation data. The calibration between JEM-X and IBIS/ISGRI can be good enough within $\sim 6\%$. In the spectral fittings, the constant factor between JEM-X and IBIS is set to be 1. The spectral
extractions for the observations of each revolution with the mean off-source angle below $\sim 5^\circ$ and IBIS detection significance level higher than $\sin 10\sigma$ are carried
out. The spectral analysis software package used is XSPEC 12.4.0x.

Generally the spectrum of accreting X-ray pulsars like 4U 2206+54 can be described by a power-law model plus a high energy exponential rolloff: $A(E)=KE^{-\Gamma}\exp(-E/E_c)$ with a photoelectric absorption below 5 keV (hereafter {\em abs*cutoffpl}). For a comparison, we also fit the spectra with other two models with the absorption: a single power law model (hereafter {\em abs*power}) and a
thermal bremsstrahlung model (hereafter {\em abs*bremss}). Then we fit the spectra of 4U 2206+54 in different revolutions with three models, which help us to justify which model is more acceptable for the spectral fitting in 4U 2206+54. In addition, we can probe the variations of spectral parameters (e.g., column density, photon index) with hard X-ray luminosity.

In Table 3, all spectral fitting parameters of three models in different revolutions are presented together. The thermal bremsstrahlung model seems to be not suitable for the spectra of the high mass X-ray binary 4U 2206+54 in most cases (the reduced $\chi^2 >2$). In addition, when the {\em abs*bremss} can fit the spectrum with the reduced $\chi^2 \sim 1$ for some revolutions, the other two models {\em abs*power} and {\em abs*cutoffpl} can also fit the spectrum well. For Rev 510 when 4U 2206+54 is brightest in our detections, the spectrum cannot be fitted by an absorbed single power-law model (the reduced $\chi^2 >4$), but an absorbed cut-off power-law model well fit the spectrum with $N_{\rm H}\sim (6.2\pm 1.3)\times 10^{22}$ cm$^{-2}$, $\Gamma\sim 1.6\pm 0.1$, $E_c\sim 32.8\pm 4.3$ keV with the reduced $\chi^2\sim 1.237$. For the other revolutions, both two models {\em abs*power} and {\em abs*cutoffpl} fit the spectrum with similar reduced $\chi^2$ values. Except for the case of Rev 510, the spectra of other revolutions fitted by the {\em abs*cutoffpl} model used in XSPEC software package give the values of $E_c$ with large error bars which may be not acceptable in statistics, then we only give the upper limits of $E_c$ values (2$\sigma$) in Table 3. Thus the cut-off power-law model may only be requested for the very bright state of 4U 2206+54. When the source becomes weaker, the single power-law model would also fit the spectrum well. In the following discussions, we only use the spectral results with the cut-off power-law model fittings.

In Fig. 6, some hard X-ray spectral samples from 3 -- 200 keV of 4U 2206+54 for eight observational revolutions were presented, in which 4U 2206+54 was in the different luminosity ranges. All spectra in Fig. 6 are fitted with the same model {\em abs*cutoffpl}.

The hard X-ray flux in the range of 5 -- 100 keV varies from $(1.5 - 7.8 )\times 10^{-10}$ erg cm$^{-1}$ s$^{-1}$ corresponding to average X-ray luminosities of $(1-8)\times 10^{35}$ erg s$^{-1}$ assuming a distance of $\sim 2.6$ kpc (Blay et
al. 2006). The spectra show the derived absorbed column density of $\sim (6 -20)\times 10^{22}$ cm$^{-2}$, and photon index range of $\Gamma\sim 1.6-2.6$. The spectral properties of 4U 2206+54 appear to change with the luminosity.

\begin{table*}

\caption{Spectral properties of 4U 2206+65 in different revolutions combined with JEM-X and IBIS. The spectra are fitted with three different models with a photoelectric absorption below 5 keV: a single power law model {\em abs*power}, a
thermal bremsstrahlung model {\em abs*bremss}, a power-law plus
exponential high energy cutoff {\em abs*cutoffpl}. The hard X-ray flux in the range of 5 -- 100 keV is presented.}

\begin{center}
\scriptsize
\begin{tabular}{l c c c c c l}

\hline \hline Rev. Num. & Model & $N_{\rm H} 10^{22}\ {\rm cm}^{-2}$    & $\Gamma$ / $kT$ (keV) & $E_{\rm c}$  &  Flux ($10^{-10}$ erg cm$^{-2}$ s$^{-1}$) & reduced $\chi^2$/$d.o.f.$ \\
\hline
510 & abs*power &  $7.8\pm 1.2$ & $2.3\pm 0.1$  & - &  $7.8\pm 0.4$  & 4.472/25  \\
  &  abs*cutoffpl & $6.2\pm 1.3$ & $1.6\pm 0.1$ & $32.8\pm 4.3$ keV & $7.4\pm 0.3$ & 1.237/24 \\
  & abs*bremss & $2.8\pm 0.8$ & $24.4\pm 0.8$ & -  &  $7.2\pm 0.3$ &  1.566/25 \\
511 & abs*power &  $18.7\pm 3.3$ & $2.5\pm 0.1$  & - &  $3.2\pm 0.3$  & 0.855/23  \\
  &  abs*cutoffpl & $16.8\pm 4.0$ & $2.4\pm 0.2$ & $<300$ keV (2$\sigma$) & $3.2\pm 0.3$ & 0.931/22 \\
  & abs*bremss & $3.5\pm 2.0$ & $22.3\pm 1.8$ &  - &  $2.9\pm 0.3$ &  2.476/23 \\
512 & abs*power &  $11.9\pm 1.7$ & $2.3\pm 0.1$  & - &  $4.1\pm 0.2$  & 1.279/21  \\
  &  abs*cutoffpl & $9.5\pm 2.1$ & $2.0\pm 0.1$ & $<230$ keV (2$\sigma$) & $4.0\pm 0.2$ & 1.118/20 \\
  & abs*bremss & $1.8\pm 1.2$ & $22.2\pm 1.2$ &  - &  $3.7\pm 0.2$ &  2.505/21 \\
513 & abs*power &  $21.7\pm 5.7$ & $2.6\pm 0.1$  & - &  $1.7\pm 0.2$  & 0.7457/22  \\
  &  abs*cutoffpl & $21.1\pm 6.8$ & $2.5\pm 0.3$ & $<500$ keV (2$\sigma$) & $1.7\pm 0.3$ & 0.7911/21 \\
  & abs*bremss & $8.9\pm 3.5$ & $17.5\pm 2.2$ &  - &  $1.5\pm 0.2$ &  1.331/22 \\
514 & abs*power &  $20.0\pm 12.1$ & $2.4\pm 0.3$  & - &  $1.8\pm 0.5$  & 0.874/17  \\
  &  abs*cutoffpl & $18.9\pm 13.8$ & $2.3\pm 0.4$ & $<140$ keV(2$\sigma$) & $1.9\pm 0.5$ & 0.923/16 \\
  & abs*bremss & $8.5\pm 7.8$ & $30.5\pm 12.2$ &  - &  $1.9\pm 0.5$ &  0.829/17 \\
516 & abs*power &  $9.7\pm 2.3$ & $2.4\pm 0.1$  & - &  $4.3\pm 0.3$  & 1.625/24  \\
  &  abs*cutoffpl & $8.7\pm 2.7$ & $2.3\pm 0.2$ & $<500$ keV(2$\sigma$) & $4.2\pm 0.3$ & 1.729/23 \\
  & abs*bremss & $1.9\pm 1.6$ & $17.5\pm 1.1$ &  - &  $3.9\pm 0.3$ &  2.758/24 \\
517 & abs*power &  $18.8\pm 10.1$ & $2.5\pm 0.3$  & - &  $1.9\pm 0.5$  & 1.101/19  \\
  &  abs*cutoffpl & $17.6\pm 11.2$ & $2.4\pm 0.4$ & $<350$ keV(2$\sigma$) & $1.9\pm 0.5$ & 0.981/18 \\
  & abs*bremss & $11.0\pm 6.8$ & $21.1\pm 10.1$ &  - &  $1.8\pm 0.5$ &  1.344/19 \\
518 & abs*power &  $19.2\pm 11.9$ & $2.5\pm 0.4$  & - &  $1.9\pm 0.4$  & 0.901/20  \\
  &  abs*cutoffpl & $18.4\pm 11.3$ & $2.4\pm 0.4$ & $<180$ keV(2$\sigma$) & $1.9\pm 0.5$ & 1.083/19 \\
  & abs*bremss & $7.9\pm 5.8$ & $19.9\pm 9.5$ &  - &  $1.7\pm 0.4$ &  1.410/20 \\
519 & abs*power &  $17.9\pm 4.1$ & $2.6\pm 0.2$  & - &  $4.6\pm 0.5$  & 0.697/24 \\
  &  abs*cutoffpl & $14.1\pm 5.0$ & $2.3\pm 0.2$ & $<200$ keV(2$\sigma$) & $4.5\pm 0.5$ & 0.639/23 \\
  & abs*bremss & $4.8\pm 2.6$ & $17.9\pm 1.8$ &  - &  $4.2\pm 0.5$ &  1.067/24 \\
520 & abs*power &  $18.9\pm 8.1$ & $2.3\pm 0.2$  & - &  $5.2\pm 0.5$  & 0.599/21 \\
  &  abs*cutoffpl & $16.8\pm 8.6$ & $1.9\pm 0.5$ & $<100$ keV(2$\sigma$) & $5.0\pm 0.5$ & 0.540/20 \\
  & abs*bremss & $12.9\pm 5.7$ & $24.8\pm 4.7$ &  - &  $4.9\pm 0.4$ &  0.568/21 \\
557 & abs*power &  $14.5\pm 4.3$ & $2.3\pm 0.2$  & - &  $1.8\pm 0.2$  & 1.263/23 \\
  &  abs*cutoffpl & $13.6\pm 5.2$ & $2.3\pm 0.3$ & $<360$ keV(2$\sigma$) & $1.8\pm 0.2$ & 1.310/22 \\
  & abs*bremss & $3.9\pm 2.9$ & $22.6\pm 3.2$ &  - &  $1.6\pm 0.2$ &  1.768/23 \\
558 & abs*power &  $17.1\pm 11.5$ & $2.4\pm 0.3$  & - &  $2.2\pm 0.6$  & 0.8671/17 \\
  &  abs*cutoffpl & $14.6\pm 12.2$ & $2.3\pm 0.5$ & $<130$ keV(2$\sigma$) & $2.1\pm 0.6$ & 0.8632/16 \\
  & abs*bremss & $8.1\pm 6.9$ & $32.2\pm 10.5$ &  - &  $1.9\pm 0.6$ &  0.8099/17 \\
559 & abs*power &  $20.1\pm 3.2$ & $2.4\pm 0.1$  & - &  $3.1\pm 0.3$  & 1.370/23 \\
  &  abs*cutoffpl & $18.5\pm 4.1$ & $2.4\pm 0.2$ & $<200$ keV(2$\sigma$) & $3.1\pm 0.3$ & 1.430/22 \\
  & abs*bremss & $5.9\pm 3.3$ & $19.8\pm 1.4$ &  - &  $2.8\pm 0.3$ &  2.178/23 \\
564 & abs*power &  $23.9\pm 6.8$ & $2.5\pm 0.2$  & - &  $2.1\pm 0.3$  & 0.695/25 \\
  &  abs*cutoffpl & $21.8\pm 8.2$ & $2.5\pm 0.3$ & $<160$ keV(2$\sigma$) & $2.1\pm 0.3$ & 0.7157/24 \\
  & abs*bremss & $9.9\pm 3.8$ & $20.6\pm 2.5$ &  - &  $1.9\pm 0.3$ &  1.125/25 \\
567 & abs*power &  $11.6\pm 2.1$ & $2.3\pm 0.1$  & - &  $3.8\pm 0.3$  & 1.008/22 \\
  &  abs*cutoffpl & $9.7\pm 2.5$ & $2.1\pm 0.1$ & $<180$ keV(2$\sigma$) & $3.7\pm 0.3$ & 0.9648/21 \\
  & abs*bremss & $1.9\pm 1.3$ & $22.4\pm 1.3$ &  - &  $3.4\pm 0.2$ &  1.989/22 \\
568 & abs*power &  $12.8\pm 2.2$ & $2.4\pm 0.1$  & - &  $3.0\pm 0.2$  & 1.474/23 \\
  &  abs*cutoffpl & $11.4\pm 2.7$ & $2.3\pm 0.1$ & $<400$ keV(2$\sigma$) & $3.0\pm 0.2$ & 1.520/22 \\
  & abs*bremss & $1.7\pm 1.5$ & $21.0\pm 1.5$ &  - &  $2.7\pm 0.2$ &  2.289/23 \\
569 & abs*power &  $14.2\pm 6.1$ & $2.5\pm 0.2$  & - &  $5.3\pm 0.9$  & 0.9242/24 \\
  &  abs*cutoffpl & $9.7\pm 7.0$ & $1.9\pm 0.4$ & $<90$ keV(2$\sigma$) & $4.9\pm 0.9$ & 0.7664/23 \\
  & abs*bremss & $2.2\pm 1.9$ & $24.3\pm 2.6$ &  - &  $4.6\pm 0.8$ &  0.899/24 \\
\hline
\end{tabular}
\end{center}

\end{table*}

To illustrate a clear description of the spectral variations with luminosity, we try to study the relations of two spectral parameters ($N_{\rm H}$, $\Gamma$) and hard X-ray flux. We have used the data points of all revolutions in Table 3 fitted with the {\em abs*cutoffpl} model. In Fig. 7, the fitted spectral parameters $\Gamma$ versus the column density and the hard X-ray flux are plotted. The photon index shows a positive correlation with the absorbed column density, $\Gamma\propto N_{\rm H}^{1.6\pm 0.2}$ with $p$-value of 2.3$\times 10^{-6}$; and an anti-correlation with the hard X-ray flux in the range of 5 -- 100 keV, $\Gamma\propto F_{\rm X}^{-1.5\pm 0.2}$ ($p$-value $\sim 1.6\times 10^{-6}$).

\subsection{Searching for cyclotron absorption line features}

\begin{figure*}
\centering
\includegraphics[angle=0,width=15cm]{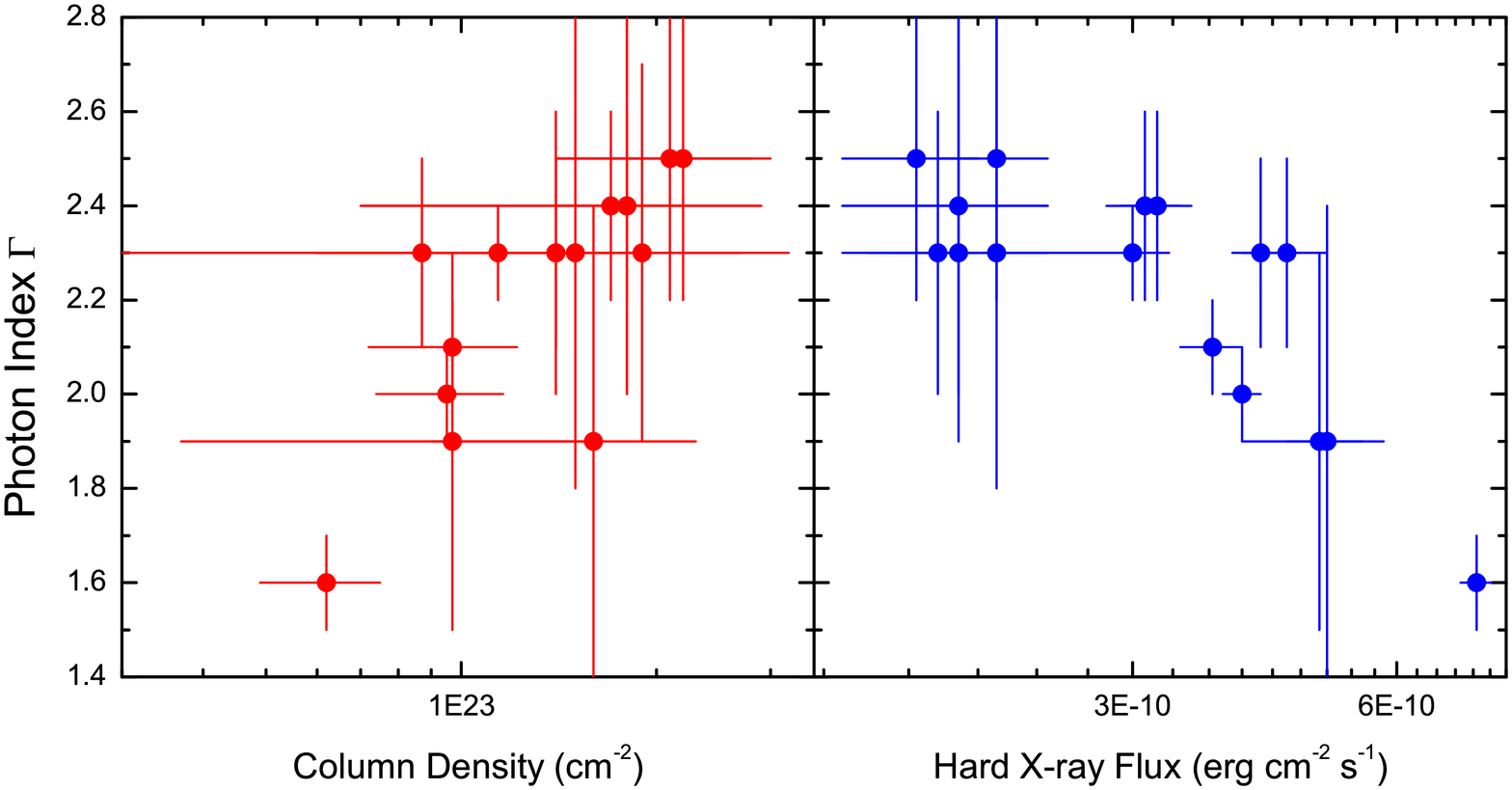}
\caption{Relations between the photon index $\Gamma$ versus the absorbed column density and $\Gamma$ versus hard X-ray flux (5 -- 100 keV) of 4U 2206+54 in different observational revolutions (data points are taken from Table 3). All spectra have been fitted with the same model of {\em abs*cutoffpl}.  }
\end{figure*}

It has been reported that there exists a possible cyclotron resonant absorption line feature around 30 keV indicated by different observations by RXTE (Torrejon et al. 2004), BeppoSAX (Masetti et al. 2004), and INTEGRAL (Blay et al. 2005; Wang 2009). The second harmonic around 60 keV was also detected in 4U 2206+54 by INTEGRAL (Wang 2009). However, this cyclotron line feature did not always appear in spectral analysis of several previous studies (Negueruela \& Reig 2001; Corbet \& Peele 2001; Torrejon et al. 2004; Wang 2010). The non-detection of cyclotron absorption lines in many experiments may still doubt on the possible existence of the magnetized neutron star in 4U 2206+54.

The production of cyclotron resonant absorption line feature near the surface of the neutron star is complicated in physics, which will sensitively depend on accretion states and accretion geometry (e.g., Araya \& Harding 1999). In addition, the fundamental line energy may also change in different accretion luminosity states and the relativistic cyclotron absorption line energy would have non-harmonic line spacing (Araya \& Harding 2000). Therefore, detection of cyclotron absorption line features is not only used to identify a magnetized neutron star in binary systems but also important for studies on accretion physics near neutron star surface.

\begin{figure}
\centering
\includegraphics[angle=-90,width=8cm]{4u2206_mean_cp.eps}
\includegraphics[angle=0,width=9.5cm]{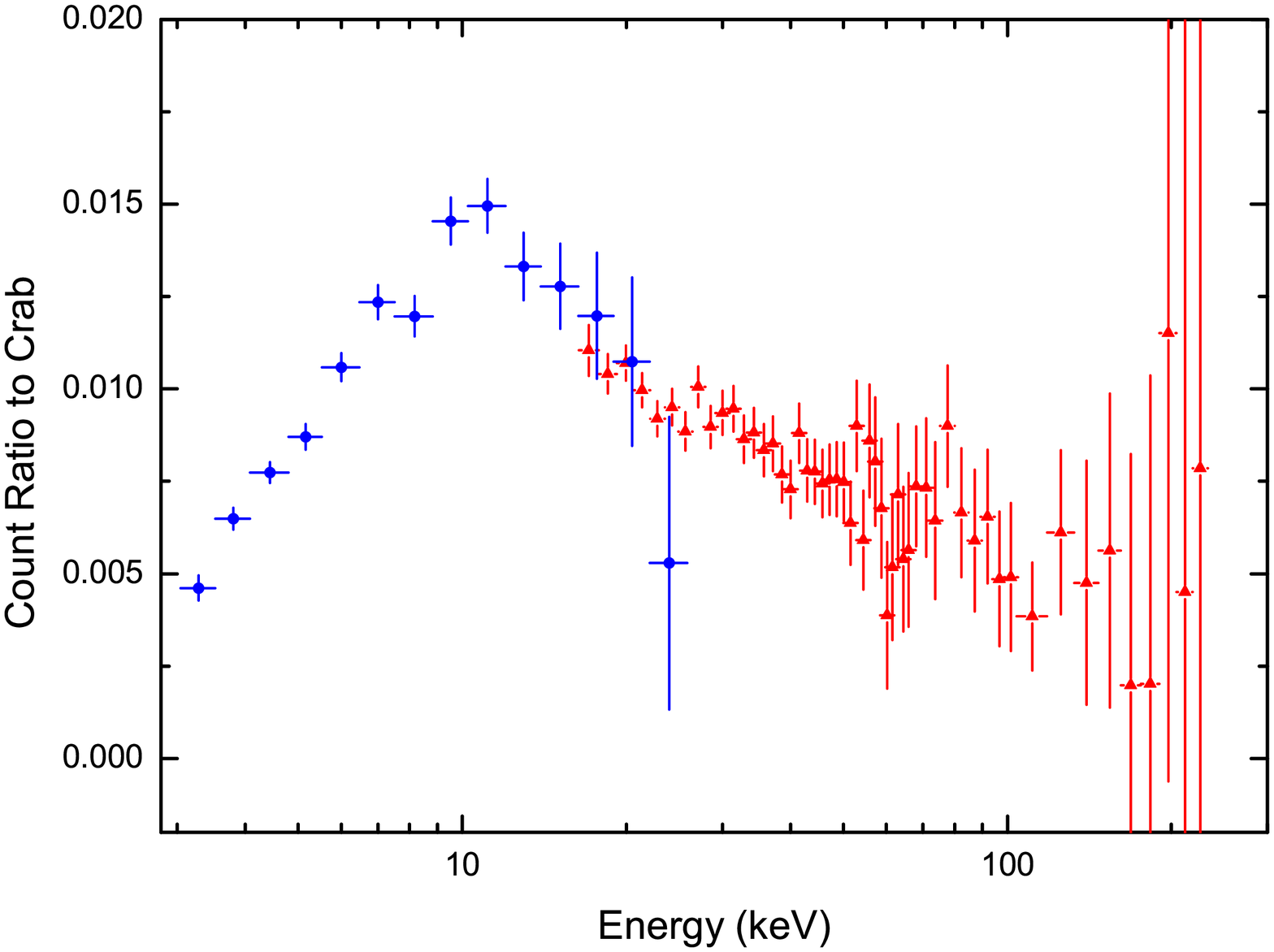}
\caption{{\bf Top:} The average hard X-ray spectrum (3 -- 200 keV) of 4U 2206+54 obtained by
the observational revolutions with the IBIS detection significance level higher than 10$\sigma$ (Table 1). The spectra are fitted with a power-law model plus high energy cutoff. {\bf Bottom:} The count ratio of the average spectrum of 4U 2206+54 to the Crab spectrum derived from Rev. 541.}
\end{figure}

Our long-term INTEGRAL observations cover the energy range of 3 -- 200 keV and different accretion states of 4U 2206+54. Then INTEGRAL observations provide us a good chance to search for the possible cyclotron absorption line features. We have done the spectral extraction in the range of 3 -- 200 keV for all observational revolutions with the IBIS detection significance level $>10\sigma$ and mean off-source angle $<5^\circ$ (Table 1) combined with the IBIS and JEM-X data. Then we check the residuals and fitted reduced $\chi^2$ values of the continuum fitting with a power-law model plus a high energy exponential rolloff on the spectra to search for the possible absorption line features (also see all fitted parameters shown in Table 2). However, because of the low S/N for one single revolution, no significant features are detected in the spectral analysis (also refer to the spectra in Fig. 6). Then we collect these data to extract an average spectrum of 4U 2206+54 from 3 -- 200 keV with a smaller energy bins (top panel of Fig. 8). The spectrum has been fitted with the {\em abs*cutoffpl} model, deriving a column density of $N_{\rm H}\sim (10.4\pm 0.9)\times 10^{22}$ cm$^{-2}$, a photon index of $\Gamma\sim 2.01\pm 0.04$ keV and $E_c\sim 73.6\pm 8.8$ keV with a reduced $\chi^2\sim 1.351$ (63 degrees of freedom). The obtained $\chi^2$ value suggests that the fitting is acceptable. The previous reported line feature around 30 keV is not confirmed in this spectral analysis. A possible structure around 60 keV is also not significant, and cannot be fitted as a possible line feature with the XSPEC package. If we try to constrain the cyclotron absorption line depth around 60 keV, we can find a upper limit of 0.8 (2$\sigma$) using the XSPEC model {\em cyclabs}.

For the further check, we also derive the count ratio of the 4U 2206+54 spectrum to the Crab one. The Crab spectrum is derived from Rev 541 which is the nearest observational revolution around the pointing observations on 4U 2206+54 in our analysis. The Crab spectrum from 3-- 200 keV can be well described by a broken power-law model of $\Gamma_1\sim 2.11\pm 0.01$, $\Gamma_2\sim 2.32\pm 0.03$ and $E_b\sim 93.2\pm 3.8$ keV, no features are discovered in the residuals. The plotted count ratio to the Crab spectrum is presented in the bottom panel of Fig. 8. No significant features are found in the spectrum ratio. Thus, the possible cyclotron absorption features around 30 keV and 60 keV are not confirmed by our analysis. Further high-sensitivity observations specially by different missions are required to check the existence of the cyclotron absorption lines in 4U 2206+54.

\section{Summary and discussion}

In this paper, we systematically study temporal and spectral properties of the slow-pulsation X-ray pulsar in 4U 2206+54 with INTEGRAL long-term monitoring observations. We confirm the existence of the slow-pulsation period around 5560 s for the neutron star in 4U 2206+54. From 2005 -- 2011, the long-term spin-down trend of the neutron star is found with the spin period evolving from $\sim 5558$ s to $\sim 5588$ s. In addition, we also search for the possible spin-down behavior of the neutron star in the short time scale. Using the phase-shift technique, we determine the spin period of the X-ray pulsar in 4U 2206+54 at $\sim 5557.1\pm 1.9$ s at MJD 54080 and discover the spin-down rate of $\dot \nu\sim -(1.8\pm 0.5)\times 10^{-14}$ Hz s$^{-1}$ from MJD 54080 -- 54260.

4U 2206+54 is a variable X-ray source, and its spectral properties also change with the X-ray luminosity. The spectrum of 4U 2206+54 can be described by an absorbed power-law model plus high energy exponential rolloff. The spectral parameters, e.g., the column density and photon index show the anti-correlation with the X-ray luminosity: higher luminosity with the lower column density and harder spectrum (smaller photon index).
We search for the possible cyclotron resonant absorption line features in the hard X-ray spectra of 4U 2206+54. The possible cyclotron absorption features at 30 and 60 keV in 4U 2206+54 are not detected with present INTEGRAL observations.

INTEGRAL observations suggest that the spectral variation pattern of 4U 2206+54 is quite similar to another very slow pulsation X-ray pulsar in 2S 0114+65. 2S 0114+65 also shows the spectral behaviors with a harder spectrum and lower absorption column density in a higher luminosity from RXTE and INTEGRAL observations (Farrell et al. 2008; Wang 2011). This spectral variation pattern in these two accretion X-ray pulsar systems suggests that they should belong to highly obscured X-ray binary systems. Anti-correlation between the hydrogen column density and X-ray luminosity has been detected in other high mass X-ray binaries like Vela X-1 (Haberl \& White 1990) and EXO 1722-363 (Thompson et al. 2007). The variations of X-ray luminosity would sensitively depend on the circumstellar environment. The increased column density and photon index during the minimum luminosity phase will be due to the absorption effect by the stellar winds. The variations of hydrogen column density with X-ray luminosity in high mass X-ray binaries have been also investigated by some authors through observations and hydrodynamic simulations (Haberl et al. 1989; Blondin et al. 1991; Blondin 1994). The stellar wind particles flowing past the compact object will be affected by different factors like gravitational, rotational and radiation pressures. The accretion wakes and possible filaments may form in the the circumstellar environment. Variability of hydrogen column density may be due to the line-of-sight motion of the neutron star near these dense materials. Though 4U 2206+54 harbors a main-sequence donor star which is quite different from the supergiant stars in 2S 0114+65, Vela X-1 and EXO 1722-363, the stellar wind properties near the compact star may have similar features, so that similar spectral variability patterns over different luminosity states are observed in these binary systems.

The neutron star of 5560-s pulsation period in 4U 2206+54 makes it as one of slowest pulsation X-ray pulsars. In the spin period - orbital period diagram for the accreting X-ray pulsars, 4U 2206+54 and the other very slow pulsation X-ray pulsar 2S 0114+65 are located in a distinct part (also see Fig. 9). In the $P_{\rm spin}- P_{\rm orbit}$ diagram, Be X-ray transients follow a positive correlation in distributions, suggesting a common formation and activity mechanisms. 4U 2206+54 and 2S 0114+65 distribute far from the Be X-ray transient line, but they may link to the supergiant wind-fed accreting binaries like Vela X-1 and EXO 1722-363 in the diagram. In observations, four X-ray pulsar systems all belong to the wind-fed direct accreting binaries, and show the similar spectral variability patterns. The donor stars in these wind-fed systems are different, supergiant stars in Vela X-1 and EXO 1722-363, a B1 supergiant star in 2S 0114+65, a main-sequence star in 4U 2206+54. In addition, 4U 2206+54 and 2S 0114+65 are the rare objects with a very slow pulsation period longer than 5000 s, and their formation channel is still unknown and quite interesting in astrophysics.

The spin period evolution of the new-born neutron star generally undergoes three states (e.g., Bhattachaya \& van den Heuvel 1991): {\em ejector state} in which neutron star spins down through the conventional spin-powered pulsar energy-loss mechanisms; {\em propeller state} in which spin period decreases by means of interaction between the neutron star magnetosphere and stellar wind of the companion; {\em accretor state} in which spin period of neutron star reaches a critical value, and the neutron star begin to accrete materials on to the surface, then switch on as the X-ray pulsar. The critical period is defined by equating the corotational radius of the neutron star to the magnetospheric radius, which gives (e.g., Pringle \& Rees 1972; Ghosh \& Lamb 1978) \beq P_{\rm cr} \simeq 18 \kappa^{3/2} {M_{\rm NS}\over 1.4M_\odot}^{-5/7}[{\mu\over 10^{30}{\rm G cm^3}}]^{6/7}[{\dot M \over 10^{15} {\rm g s^{-1}}}]^{-3/7} {\rm s}, \enq where $\kappa\sim 0.5-1$ is the geometrical parameter of the accretion flow, $\mu$ is the dipole magnetic moment of the neutron star, and $\dot M=\pi r_{\rm G}^2\rho_\infty V_{\rm rel}$ is the mass with which a neutron star interacts in a unit time as it moves through the stellar wind of the density $\rho_\infty$ with a velocity $V_{\rm rel}=\sqrt{V_{\rm NS}^2+V_{\rm w}^2}$, $r_{\rm G}=2GM_{\rm NS}/V_{\rm rel}^2$ is the Bondi radius. According to Eq. (4), the spin period of the neutron star could increase during the {\em ejector} and {\em propeller} states to the longest period of several hundred seconds but less than $\sim$ 1000 s for the neutron star of magnetic field $B<B_{\rm cr}=4.4\times 10^{13}$ G (see e.g. Urpin et al. 1998).

However, the superlong spin period ($>1000$ s) neutron star in high mass X-ray binary cannot be formed in this standard scenario. Some theorists have provided other possible mechanisms. Li \& van den Heuvel (1999) have suggested that a slow spin period is possible if the neutron star
was born as a magnetar with an initial magnetic field $\geq
10^{14}$ G, allowing the neutron star to spin down to the spin period range longer than 1000 s during the canonical {\em propeller} phase within the lifetime (Myr) of the companion. They also suggested that this ultra-strong magnetic field could decay to the normal value ranges of $10^{12}-10^{13}$ G within a few million years, so that 4U 2206+54 and 2S 0114+65 may be defined as magnetar descendants. The alternative suggestion for the origin of long-period X-ray pulsars proposed by Ikhsanov (2007) shows that an additional evolution phase {\em subsonic propeller} state between the transition from known {\em supersonic propeller} state to {\em accretor} state could allow the spin period increases up to several thousand seconds which may not require the assumption of magnetars, which is the so-called break period given by Ikhsanov (2001): \beq  P_{\rm br} \simeq 2000 {M_{\rm NS}\over 1.4M_\odot}^{-4/21}[{B_{\rm surf}\over 0.3B_{\rm cr}}]^{16/21}[{\dot M \over 10^{15} {\rm g s^{-1}}}]^{-5/7} {\rm s}, \enq where $B_{\rm surf}$ is the surface magnetic field of the neutron star.

\begin{figure}
\centering
\includegraphics[angle=0,width=9.5cm]{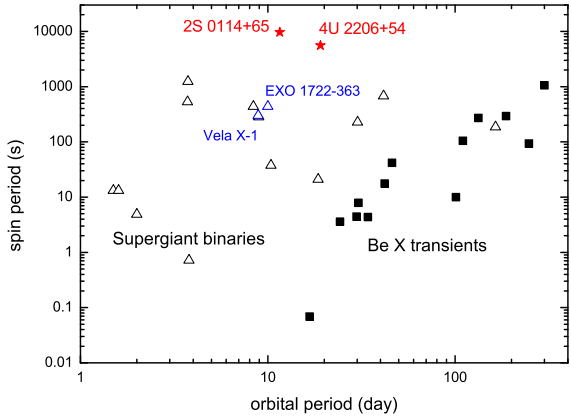}
\caption{The spin period - orbital period diagram for the accreting X-ray pulsars in high mass X-ray binaries. The data point for 2S 0114+65 is taken from Wang (2011) and Farrell
et al. (2008); the data point for 4U 2206+54 is taken from this work and Wang
(2009); the data point of EXO 1722-363 is taken from Thompson et al (2007); the data points of other supergiant binaries (triangles) including Vela X-1 and Be X-ray transients (squares) are taken from Bildsten et al. (1997), Sidoli (2011) and Haberl et al. (2012). There
exists a positive correlation between $P_{\rm spin}- P_{\rm orbit}$
for the Be transient systems which have the transient accretion disk during outbursts. Two very slow pulsation X-ray pulsars 2S 0114+65 and 4U 2206+54, and the supergiant X-ray binaries EXO 1722-363 and Vela X-1 all belong to wind-fed direct accretion systems and emit persistent and variable X-rays. If the possible relation between $P_{\rm spin}- P_{\rm orbit}$ exists in the wind-fed systems, it should be very different from that of the Be transients. }
\end{figure}

\begin{figure}
\centering
\includegraphics[width=9.5cm]{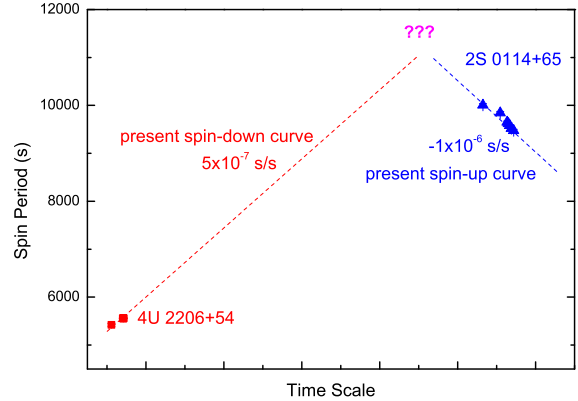}
\caption{The imaginary picture which shows the spin period evolution history of two very slow pulsation neutron stars in high mass X-ray binaries 4U 2206+54 and 2S 0114+65. The spin period data points (red squares) of 4U 2206+54 are taken from Reig et al. (2009), Finger et al. (2010) and this work. The spin period data points (blue triangles) of 2S 0114+65 are taken from Finley et al. (1992), Hall et al. (2000), Farrell et al. (2008) and Wang (2011). The dashed lines show their possible spin evolution channels. 4U 2206+54 undergoes a spin-down process with a rate of $\sim 5\times 10^{-7}$ s s$^{-1}$ (from Finger et al. 2010; Reig et al. 2012 and this work), which will make the spin period of 4U 2206+54 longer than 10000 s within 300 years if the rate is stable in next a few hundred years. While, 2S 0114+65 undergoes a fast spin-up process in the last twenty years with spin period variations from $\sim$ 10000 s to present 9500 s (Wang 2011). It is possible that there exists a candidate neutron star binary with a spin period much longer than 10000 s which may be the product of long-term spin down of the 4U 2206+54-like neutron star and would undergo the spin-up transition to form the 2S 0114+65-like source. }

\end{figure}

The different measurements (Figs. 3 \& 4) showed the spin-down trend for the neutron star in 4U 2206+54 in the last twenty years with an average spin-down rate of $\sim 5\times 10^{-7}$ s s$^{-1}$. According to the standard evolutionary scenario, the maximum spin-down rate in the accretor stage is given by $\dot P\sim 2\pi B^2R^6_{\rm NS}/(GMI)$, where $I$ is the momentum of inertia. This equation implies $B> 10^{14}$ G for 4U 2206+54. Recently, a new theory of quasi-spherical accretion for X-ray pulsars is developed (Shakura et al.\ 2012), the magnetic field in wind-fed neutron star systems is given by \beq B_{14}\sim 70\dot M^{1/3}_{15}V^{-11/3}_{300}({P_{1000}\over P_{orb}})^{11/12} G. \enq This also gives the ultrastrong magnetic field of $>10^{14}$ G for the case of 4U 2206+54 if we take the orbital period as 19 days.

Then the discovered long-term spin-down trend interpreted that the neutron star in 4U 2206+54 should be an accreting magnetar with a magnetic field of $> 10^{14}$ G. If so, the system of 4U 2206+54 should be very young, and the magnetic field of the new-born neutron star could be even higher due to the field decay. In addition, magnetar activity would emit X-rays with a typical luminosity of $\sim 10^{35}-10^{36}$ erg s$^{-1}$ and a typical power-law spectrum in hard X-ray bands (10 -- 100 keV) of photon index $\Gamma\sim 1-1.5$ for known magnetar candidates (see Kuiper et al. 2004, 2006; Wang 2008). The observed spectrum of 4U 2206+54 in hard X-rays is still the typical accretion-powered pulsar spectrum, power-law with a high energy cutoff, and even in the relatively weak states, the spectrum was fitted with a single power-law model of $\Gamma\sim 2.5$ (also see Wang 2009), quite different from the known magnetar candidates. The persistent X-ray luminosity of magnetars is relatively stable, but the X-ray luminosity of 4U 2206+54 varies from $10^{34} -10^{36}$ erg s$^{-1}$ from long-term X-ray observations (Torrejon et al. 2004; Masetti et al. 2004; Blay et al. 2005; Corbet et al. 2007; Wang 2009, 2010). These inconsistences must lead to the assumption that present magnetar-activity powered X-ray luminosity is lower than $10^{34}$ erg s$^{-1}$ for the case of 4U 2206+54.

We could not exclude the possibility of the magnetar nature in 4U 2206+54, but some more complicated issues need be addressed in further studies. 4U 2206+54 could be a magnetized neutron star with a normal magnetic field of 3$\times 10^{12}$ G assuming the electron cyclotron line feature round 30 keV (Blay et al. 2005; Wang 2009), which is similar to other accreting pulsars in high-mass X-ray binaries. However, the problem is that the neutron star is difficult to spin down to more than 5000 s in the present accretion theory, except for the magnetar assumption or very low accretion rate ($\sim 10^{12}$ g s$^{-1}$, see Fig. 1 of Li \& van den Heuvel 1999). In the mean observed X-ray luminosity range of $10^{35}-10^{36}$ erg s$^{-1}$ in the range of 3 -- 100 keV, the derived mass accretion rate is at least about $5\times 10^{14}-5\times 10^{15}$ g s$^{-1}$ for 4U 2206+54. The difficulty and uncertainties in explaining the long spin period still exist. It is possible that the X-ray pulsar in 4U 2206+54 may not follow the present standard evolution models in close
binaries.

The very slow pulsation neutron star in 2S 0114+65 is undergoing the spin-up process from more than 10000 s to the present 9500 s with a spin-up rate of $\sim 10^{-6}$ s s$^{-1}$ in the last twenty years (see Wang 2011). If the X-ray pulsar in 2S 0114+65 continues to spin up, 2S 0114+65 would reach its possible equilibrium period less than $\sim$ 1000 s in next a few hundred years. The spin period evolution of the neutron star in 4U 2206+54 is also detected with INTEGRAL observations (Fig. 4). With comparing different observations in the last twenty years (Fig. 3), we still confirm that the neutron star in 4U 2206+54 is undergoing a spin-down trend with a mean spin-down rate of $\sim 5\times 10^{-7}$ s s$^{-1}$ (also see Finger et al. 2010; Reig et al. 2012). If so, with the present rate, 4U 2206+54 would spin down to a rotation period longer than 10000 s within $\sim 300$ years (also see Fig. 10). There exists a possible link between the two slow-pulsation X-ray pulsars from the cartoon picture in Fig. 10. 4U 2206+54 could spin down to a possible slowest pulsation neutron star with spin period much longer than 10000 s, and then undergo a transition to the spin-up channel, becoming the 2S 0114+65-like system. The evolution channel from the 4U 2206+54-like system to the 2S 0114+65-like system could be supported by the their companion types: according to the evolutional tracks of massive stars (e.g., Meynet et al. 1994), the possible progenitor of a B1 supergiant in 2S 0114+65 is an O9.5 V star consistent with the main-sequence star type in 4U 2206+54. Some other supergiant X-ray binaries like Vela X-1 and EXO 1722-363 may be the product of 2S 0114+65-like sources which evolves to the equilibrium period less than $\sim 1000$ s through spin-up processes. According to the present spin-down and spin-up rates of the two very slow pulsation neutron stars, the total lifetime of the slow pulsation X-ray pulsar phase is very short with a time scale of $\sim 10^3$ yrs, so the object is very rare but fortunately we have detected at least two very slow pulsation neutron star systems.

We have suggested an evolution track for the slow-pulsation neutron stars in wind-fed accreting binaries. There may exist the physical link among these slow-pulsation X-ray pulsars: 4U 2206+54 and 2S 0114+65 are the younger systems undergoing rapid spin period evolutions, from the spin-down phase of 4U 2206+54 to the spin-up phase of 2S 0114+65; supergiant X-ray binaries, like Vela X-1 and EXO 1722-363 are the older systems in the equilibrium spin period range after the rapid evolution phase. In the equilibrium, accreting X-ray pulsars in wind-fed systems would not undergo long-term spin-up/spin-down evolution trends, the torque changes and fluctuations during accretion will occur which has been observed in Vela X-1 and some other supergiant neutron star binaries (see Bildsten et al. 1997). This evolution track assumption requires further long-term observations, specially monitoring the spin-down trend of 4U 2206+54: the spin-down process continues in a longer time scale or there exist torque fluctuations or even the torque changes in future. These important issues will help us to well understand the origin and evolution of very slow pulsation neutron star systems, and provide new information on the evolution channels of accreting neutron star high mass X-ray binaries, specially these accreting magnetar candidates.

\section*{Acknowledgments}
This paper is based on observations of
INTEGRAL, an ESA project with instrument and science data center
funded by ESA member states. The author is grateful to the referee for the fruitful suggestions to improve the manuscript and also thanks MPE for the help in the data processing. The work is supported by the
National Natural Science Foundation of China under grant 11073030.

\end{document}